\documentclass[twocolumn,showpacs,preprintnumbers,amsmath,amssymb]{revtex4}

\usepackage{graphicx}
\usepackage{dcolumn}
\usepackage{bm}

\newcommand {\slsh} [1] {\not{\hbox{\kern-2pt${#1}$}}}

\def\drawbox#1#2{\hrule height#2pt
         \hbox{\vrule width#2pt height#1pt \kern#1pt
               \vrule width#2pt}
               \hrule height#2pt}

\def\Asym#1#2{\vcenter{\vbox{\drawbox{#1}{#2}
               \kern-#2pt       
               \drawbox{#1}{#2}}}}

\def\asymm{\Asym{6.4}{0.3}}

\def\basymm{\overline{\asymm}}

\newcommand{\None}{${\cal N}=1\ $}

\begin{document}

\preprint{CERN-TH/2003-151, FTPI-MINN-03/18, UMN-TH-2207/03}

\title{SUSY Relics in  One-Flavor QCD from \\ a New 1/N Expansion}

\author{A. Armoni}
\email{adi.armoni@cern.ch}
\author{M. Shifman}
\email{michael.shifman@cern.ch}
\altaffiliation[Permanent address:]{ William I. Fine Theoretical Physics Institute, 
University of Minnesota, Minneapolis, MN 55455, USA}
\author{G. Veneziano}
\email{gabriele.veneziano@cern.ch}
\affiliation{Theory Division, CERN \\CH-1211 Geneva 23, Switzerland}

\date{\today}

\begin{abstract}
We suggest a new large $N_c$ limit for multi-flavor QCD.
Since  fundamental and
two-index antisymmetric representations are equivalent in SU(3), we have 
the
option to define SU$(N_c)$ QCD
keeping quarks in the latter. We can then define a new $1/N_c$
expansion  (at fixed number of flavors $N_f$) that shares appealing
properties with the topological (fixed $N_f/N_c$) expansion while
being more suitable for theoretical analysis.  In particular, for $N_f
=1$, our large-$N_c$ limit gives a theory that we recently proved to
be equivalent, in the bosonic sector, to \None supersymmetric
gluodynamics. Using known
properties of the latter, we derive several  qualitative and
semi-quantitative predictions for $N_f=1$ massless QCD that can be easily
tested in lattice simulations. Finally, we comment on possible
applications for pure SU(3) Yang-Mills theory and real QCD.
\end{abstract}


\maketitle

\noindent
 Very few techniques are
available for analytical studies of non-perturbative properties of
non-supersymmetric gauge theories such as QCD. Among the
most promising ones, large-$N$ expansions play a special role,
in particular  because of  their conjectured connection to string 
theories.

The simplest and the oldest $1/N_c$
expansion in QCD is that suggested  by 't Hooft \cite{tHooft}.
The 't Hooft limit assumes $N_c \rightarrow \infty$ while keeping the
 't Hooft coupling
$g^2 N_c$ fixed. If  the number of quark flavors
is fixed too (i.e. does not scale with $N_c$)
then each quark loop is suppressed by $1/N_c$.  Only quenched planar
diagrams survive in the  leading order.
Non-planar diagrams with ``handles'' are suppressed by $1/N_c^2$ per
 handle. Thus, the corrections to the leading approximation run in powers of
$(N_f/N_c)$ and $1/N_c^2$. The 't Hooft expansion led to a number of 
notable successes
in such issues as the Zweig rule, the $\eta '$ mass formula (see below), 
and so on.
Unfortunately, nobody  succeeded in fully solving QCD even to the 
leading order in
the  't~Hooft expansion.

In the range of questions where the quark loops are important,
a better approximation is provided by the
topological expansion (TE) \cite{TE}  in which $N_f/N_c$ is  kept
fixed in the large-$N_c$ limit, rather than $N_f$. Then, in the leading
order,  TE preserves {\em all} planar diagrams, including quark loops.
This is
easily seen by  slightly  modifying   \cite{TE} the  't Hooft 
double-line notation --- adding
a flavor line  to the single color line for quarks. In the leading 
(planar) diagrams the
quark loops are ``empty" inside, since gluons do not attach to the 
flavor line.
Needless to say,  obtaining analytic results in   TE is even harder  
than in
the  't Hooft case.

In this letter we  propose a new
large-$N_c$ expansion that  shares some   advantages of   TE  while
retaining  a significant predictive power.
The results obtained in the new large-$N_c$ limit
are complementary to those derived in the 't Hooft limit.

Our basic idea is as follows. Let us start from $N_c=3$ QCD with $N_f$
quark flavors, which may or may not be massless. (For definiteness
we will consider the massless case.)
The quark can be described by a Dirac field transforming in the
fundamental representation of SU(3)$_{\rm color}$, or, {\em 
equivalently}, in
the two-index antisymmetric representation (plus their complex 
conjugates).
In extrapolating from $N_c=3$ to arbitrary $N_c$, the former alternative
leads to the 't Hooft limit. We will explore the latter alternative,
representing the quark of a given flavor by  a Dirac field in the  {\em 
two-index antisymmetric} representation. Now,
taking the limit $N_c \rightarrow \infty$ at $g^2N_c$ and $N_f$ fixed
does not decouple the quark loops since, for large $N_c$, the number of
degrees of freedom in the anti-symmetric field scales as $N_c^2$. This
is the starting element of  our new  $1/N_c$ expansion.
For reasons explained in  \cite{Armoni:2003gp},
it will be referred to as the orientifold large-$N_c$ limit.
The leading order of this new expansion corresponds to
the sum of all planar diagrams, in the same way as in TE,
but with the crucial difference that quark loops are now ``filled",
because the second line in the fermion propagator
is now also a color line.

The orientifold large-$N_c$ limit is, therefore, unquenched,
and its 't Hooft diagrams  look precisely as those of QCD with $N_f$ 
Majorana
fields  in the adjoint representation, modulo  reversal of
some arrows in the fermion loops.
Let us call the SU($N_c$) Yang--Mills theory with $N_f$  Majorana
fields in the adjoint representation
{\em adjoint QCD}. Trivially extending  our original arguments
  \cite{Armoni:2003gp}, one can actually prove \cite{Armoni:2003jk}
that the orientifold  large-$N_c$ limit of QCD is equivalent, in its 
bosonic sector,
to the adjoint QCD just mentioned. This can be shown to be the case
even in the presence of quark masses.

Adjoint QCD can be seen, in turn, as a softly-broken version of
supersymmetric Yang--Mills theory (SYM)
with $N_f -1$ additional adjoint chiral superfields.
This class of models is currently under intense scrutiny
in connection with the works of Refs. \cite{DV,SW}.
Soft-breaking mass terms are ascribed to
the scalar fields, so that
the scalar sector decouples.
We will briefly comment on the $N_f >1$ case later,
focusing now
on the particular case  $N_f =1$.

One-flavor QCD in the orientifold large-$N_c$ limit
becomes SU($N_c$) Yang--Mills theory with one Dirac
spinor in the antisymmetric representation.
As was shown in Ref.~\cite{Armoni:2003gp},
this theory is planar-equivalent to adjoint QCD with
$N_f=1$, which is nothing but SU($N_c$) super Yang--Mills theory.
This result was a development of  the previously
formulated Strassler's conjecture \cite{Strassler:2001fs} and studies
of brane configurations of type 0 string theory \cite{type0}.
The planar equivalence between \None SYM theory
and orientifold field theories amounts to the following statement:
the SU($N_c$) gauge theory with two Weyl
fermions in the two-index antisymmetric representation,
$$\asymm +
\basymm$$
(i.e. one Dirac anti-symmetric fermion)
  at  $N_c\to\infty$ is equivalent, in a bosonic subsector, to
  \None gluodynamics. Not only were
the planar  \None SYM theory and the planar
orientifold field theory prove
to be equivalent at the perturbative level: we argued
that the full (perturbative and non-perturbative) partition functions
of the two theories become equivalent after integration over the 
respective
fermions, at the planar level.

It is worth noting that the two theories are not {\em fully} identical.
In particular the color-singlet spectrum of
the orientifold theory consists  only of bosons and does not
include composite fermions at $N_c\to\infty$. Some
consequences of the (partial) equivalence were
discussed in detail in \cite{Armoni:2003gp,Armoni:2003jk}.

Now we will use the above planar equivalence to make predictions
for  one-flavor QCD, keeping in mind that they are expected to be
valid up to corrections of the order of $1/N_c =1/3$ (barring
large numerical coefficients):

(i) Confinement with a mass gap. Here we assume that large-$N_c$ \None
gluodynamics is a confining theory with a mass gap. Alternatively,
if we start from the statement that one-flavor QCD confines,
we arrive at the statement that so does \None SYM theory,
while the mass gaps are dynamically generated in both theories.

(ii) Degeneracy in the color-singlet bosonic
spectrum. Even/odd parity mesons (typically mixtures of fermionic and 
gluonic color-singlet
  states) are expected to be degenerate. In particular,
\begin{equation}
{m^2 _ {\eta'} \over m^2 _ \sigma}=1 + O(1/N_c)
\,, \qquad\mbox{one-flavor QCD}\,,
\label{etasigma}
\end{equation}
where $\eta '$ and $\sigma$ stand for $0^-$ and $0^+$
mesons, respectively.
This follows from the exact degeneracy in \None SYM theory. Note that the
$\sigma$ meson is stable in this theory, as there are no pions.
The prediction \eqref{etasigma} should be taken with care, (i.e.
a rather large numerical coefficient in front of $1/N_c$ may occur), since 
the
$\eta'$ mass is given by the anomaly (the WV
formula \cite{Witten:1979vv,Veneziano:1979ec}), whereas the $\sigma$
mass is more ``dynamical.'' The
degeneracy among the even/odd parity mesons should improve at higher
levels of the expected Regge trajectory.

(iii) Bifermion condensate. \None SU$(N_c)$ gluodynamics has a
bifermion condensate  \cite{gluinocond} that can take $N_c$ distinct 
values:
\begin{eqnarray}
& & \langle \lambda  \lambda \rangle_k  \sim M_{\rm uv}^3\,  e^{- \tau 
/N_c } e^{2 i \pi k/N_c} =
c \Lambda^3 \, e^{i(\theta +2 \pi k)  /N_c} \, , \nonumber \\
& &  \,\,\,\; k = 0, 1, 
\dots, N_c-1 ,
\label{bfcsusy}
\end{eqnarray}
with
$$\tau =\frac{8\pi^2}{g^2} - i\theta\,$$
and $c$ a calculable numerical coefficient.
  The finite-$N_c$ orientifold field theory is  non-supersymmetric,
and here we expect (taking account of pre-asymptotic $1/N_c$
corrections) $N_c-2$ degenerate vacua with
\begin{eqnarray}
& & \langle \bar \Psi _L \Psi _R\rangle_{k'}  \sim M_{\rm uv}^3\,  
\exp\left\{-
{8\pi ^2\over
g^2 (N_c+4/9)} + i{ \theta + 2 \pi k'\over {N_c-2}}\right\} 
\nonumber \\
& & \sim 
c' \Lambda^3 \, \exp\left\{i{\theta + 2 \pi k'\over N_c-2}\right\}
\, , \nonumber \\ 
& & \quad k' = 0,1, \dots , N_c-3\,.
\label{bfco}
\end{eqnarray}
The term $4/9$ in \eqref{bfco} is due to the one-loop
$\beta$ function of the orientifold
field theory,
$b=3N_c +{4\over 3}$, while $N_c-2$ in \eqref{bfco} is twice the dual
Coxeter number of the
antisymmetric representation (fixing the coefficient of the axial 
anomaly),
$${\rm tr}\,\, T_{\asymm} ^a\,\, T_{\asymm} ^b = {1\over2}(N_c-2)\,
  \delta ^{ab}\,.
$$
Finally, $c'$ is a normalization factor. For a suitably normalized
renormalization-group-invariant quark bilinear we predict
\begin{equation}
\langle \bar \Psi _L \Psi _R\rangle = c'
\Lambda^3 \, e^{i\theta} \,, \qquad c' = c(1 + O(1/N_c)) \,,
\label{bfQCD}
\end{equation}
i.e.  a non-vanishing condensate and a single true vacuum. Note that the
naive prediction of three degenerate vacua from the \None SYM
theory is useless
because in this case the $1/N_c$ correction is large ($N_c \rightarrow
N_c-2$ in passing from \eqref{bfcsusy} to \eqref{bfco}).
We  expect, however, the absolute value of the condensate
to be in reasonable agreement with the supersymmetry-based prediction.

(iv) The accuracy of ``dynamical'' predictions, namely those results that
are not saturated by one loop (i.e. the anomalies), can be inferred from
perturbative arguments. In  one-flavor QCD
the first coefficient of the $\beta$ function is $b= 31/3$,
while in  adjoint QCD with $N_f=1$ it becomes $b= 27/3$.

In fact, in the very same approximation it can be asserted that
the $\beta$ function of the one-flavor QCD should coincide
with the exact  NSVZ beta function \cite{nsvzbeta}:
\begin{equation}
\beta = -{1\over 2\pi}\,  {{9 \alpha ^2} \over {1- 3\alpha / 2\pi}}\, .
\end{equation}
  Thus, for the (relative) value of the 2-loop $\beta$-function
coefficient, we predict $+3 \alpha / 2\pi$, to be compared
with  the exact
value in one-flavor QCD:
$$+ \frac{134}{31}\frac{ \alpha }{2\pi} \approx 4.32 \,\frac{\alpha }
{ 2\pi}\,.$$
  Note that our $1/N_c$ expansion, unlike 't Hooft's, over-emphasizes
the quark-loop contributions, and thus makes the theory less 
asymptotically free
than in reality--- the  opposite of what happens in 't Hooft's 
expansion.
Parametrically, the error in our way of
treating one-flavor QCD  is  $1/N_c$ rather than  $1/N^2_c$. This  is 
because
there are $N^2_c-1$ gluons and $N^2_c-N_c$ fermions in the orientifold 
field
theory.

Now we will comment on possible applications for the pure SU(3)
Yang--Mills theory. As was explained, one-flavor QCD is approximated (to
$O(1/N_c)$)  by a supersymmetric theory. On the other hand, a different
$1/N_c$ expansion ---  the 't Hooft one  --- connects one-flavor QCD  to 
the
pure SU(3) gauge theory. Therefore, the pure SU(3) theory is also
approximated, in a sense,  by a supersymmetric theory.
Although we suspect that in this case
the two approximations accumulate, so that errors are large,
it is reasonable to ask whether there are relics of SUSY in
the SU(3) Yang-Mills theory.

In fact, it has been known for a long time \cite{Novikov:xj}
that such relics do exist in pure gauge theory, although at that time
they were not interpreted in terms of supersymmetry.
Indeed, it was shown  \cite{Novikov:xj}
that one can use an approximate holomorphic formula\, 
\begin{equation}
\langle {\rm tr}\, F^2 + i\,{\rm tr}\tilde F F \rangle =
M^4_{\rm uv}\, e^{-\tau / 3}\,.
\label{hf}
\end{equation}
(For this equation to be holomorphic in $\tau$,
renormalization-group-invariant and,
in addition, to  have the correct $\theta$ dependence,
we must use $b=12$ rather than the actual value $b=11$ in SU(3) 
Yang--Mills theory. The decomposition $11=12-1$ has a deep physical
meaning: $-1$ presents a unitary contribution in the one-loop $\beta$
function, while $12$ is a {\em bona fide} antiscreening, see e.g. \cite{Netal}).
The holomorphic dependence is a relic of supersymmetry.
Equation (\ref{hf}) implies, in particular,
that the topological susceptibility $\chi$
is expressible in terms of the expectation value of
${\rm tr}\,F^2$:
\begin{eqnarray}
12 \chi & \equiv & -i\int\, d^4x\,\left\langle \frac{\sqrt 3}{16\pi^2}\,
F^a \tilde{F} ^a (x)\,,
\,\, \frac{\sqrt 3}{16\pi^2}\, F ^a \tilde{F} ^a(0)
\right\rangle_{{\rm conn}} \nonumber \\ 
 & = & \left\langle\frac{1}{8\pi^2 } F^a F^a 
\right\rangle \, ,
\label{twopoint}
\end{eqnarray}
  The numerical value of the left-hand side is known, either from the
WV formula \cite{Witten:1979vv,Veneziano:1979ec}
or from lattice measurements \cite{Alles:1996nm}, to be
$\approx 1.3\times 10^{-2}\,\,{\rm GeV}^4$. The gluon condensate
on the right-hand side is that of  pure Yang--Mills theory and
 was estimated to be approximately twice larger  (see
  Ref.~\cite{Novikov:xj}, Sect. 14)
than the gluon condensate in actual QCD, $$\left \langle {1\over 4\pi^2} F^a
  F^a \right \rangle _{\rm QCD} \approx 0.012 \, {\rm GeV}^4,$$ see
  Ref. \cite{SVZ}. If so, the numerical value of the right-hand
side is $\approx 1.2\times 10^{-2}\,\,{\rm GeV}^4$. Note that the
  phenomenological estimate of the gluon condensate and the factor 2
  enhancement mentioned above are valid up to $\sim 30\%$. 

Finally, we want to comment on {\em real QCD}. Let us assume for a moment
that the $\sigma$ mass is not very sensitive to the number of flavors.
On the other hand, according to the WV formula
\cite{Witten:1979vv,Veneziano:1979ec} and neglecting quark masses, the
$\eta'$ mass scales like $\sqrt{N_f}$; we can therefore
 extrapolate the relation \eqref{etasigma} to
obtain a prediction for real QCD
\begin{equation}
m_{\eta '} \sim \sqrt 3 m_{\sigma}.
\end{equation}
  Although  $\sigma$ is very broad in real QCD, it is amusing that the above
relation is indeed in qualitative  agreement with the position of the
enhancement in the $\pi\pi$ channel.

The connection between  {\em three}-flavor QCD
and {\em adjoint} QCD with $N_f=3$ can be used to relate
the patterns of the chiral symmetry breaking
in these two theories, see Ref.~\cite{Armoni:2003jk}.

In 't Hooft's large $N_c$ limit 
the width of $q \bar{q}$ mesons scales as $1/N_c$,
as compared to $1/N_c^2$ for glueballs, i.e.
the latter are expected to be relatively more narrow. This is quite weird, given that no
glueball was identified so far, in spite of decades of searches.

Our $N_c$ counting, instead, predicts that the widths of both
quarkonic and gluonic mesons scale as $1/N_c^2$.
Given that glueballs are expected to be heavier
than their quark counterparts, this might explain
why  many narrow quarkonia have been found but no comparably-narrow
glueballs.

In conclusion, we have argued that the orientifold theory generalization 
of  QCD to arbitrary $N_c$
  can be used (through its  SYM limit) in order to make predictions for  
one-flavor QCD.
  Hopefully, these will be tested in
lattice simulations in the near future.

{\it Note added}
Meanwhile, the calculation of the quark condensate in one-flavor QCD,
starting from SUSY gluodynamics, as outlined in the text,
has been completed \cite{ASV3} and has exhibited a remarkable
agreement with the currently accepted ``phenomenological"
value.

\begin{acknowledgments}

We would like to thank J. Barbon, P. H. Damgaard, P. de Forcrand, 
P. Di Vecchia, L. Giusti,
N. Itzhaki, B. Lucini, M. L\"{u}scher, F. Sannino, A. Schwimmer, and M. Teper,
for discussions. The work of M.S. is supported in
part  by DOE grant DE-FG02-94ER408.
\end{acknowledgments}

\end{document}